\documentclass[11pt]{article}
\usepackage{amsmath}
\usepackage{amssymb}

\usepackage{natbib}
\usepackage{graphicx}

\usepackage{color} 

\usepackage[hmargin=1.8cm, top=1.8cm, bottom=2.5cm]{geometry}
\usepackage[labelfont=bf,labelsep=period, footnotesize]{caption}
\usepackage{sidecap}

\date{}

\usepackage{cite}

\begin{document}

\begin{flushleft}
{\LARGE
\textbf{Compensatory evolution and the origins of innovations}
}
\bigskip
\\
Etienne Rajon$^{1,2,3}$, 
Joanna Masel$^{1}$
\\
\bigskip
\bigskip
$^1$ Department of Ecology and Evolutionary Biology, University of Arizona, Tucson, AZ 85721, USA
\\
$^2$ Department of Biology, University of Pennsylvania, Philadelphia, PA 19104, USA
\\
$^3$ E-mail: rajon@sas.upenn.edu\\
\end{flushleft}

\vspace{0.5cm}

\begin{abstract} Cryptic genetic sequences have attenuated effects on phenotypes. In the classic view, relaxed selection allows cryptic genetic diversity to build up across individuals in a population, providing alleles that may later contribute to adaptation when co-opted -- \textit{e.g.} following a mutation increasing expression from a low, attenuated baseline. This view is described, for example, by the metaphor of the spread of a population across a neutral network in genotype space. As an alternative view, consider the fact that most phenotypic traits are affected by multiple sequences, including cryptic ones. Even in a strictly clonal population, the co-option of cryptic sequences at different loci may have different phenotypic effects and offer the population multiple adaptive possibilities. Here, we model the evolution of quantitative phenotypic characters encoded by cryptic sequences, and compare the relative contributions of genetic diversity and of variation across sites to the phenotypic potential of a population. We show that most of the phenotypic variation accessible through co-option would exist even in populations with no polymorphism. This is made possible by a history of compensatory evolution, whereby the phenotypic effect of a cryptic mutation at one site was balanced by mutations elsewhere in the genome, leading to a diversity of cryptic effect sizes across sites rather than across individuals. Cryptic sequences might accelerate adaptation and facilitate large phenotypic changes even in the absence of genetic diversity, as traditionally defined in terms of alternative alleles.
\end{abstract}

\vspace{1cm}

\section*{Introduction}
Populations that contain or produce a greater range of heritable phenotypic variants are more likely to adapt to novel environments. However, ``new'' phenotypes often have low fitness in the ancestral environment, limiting populations' ability to accumulate potentially adaptive genetic diversity. This problem is partly resolved when genetic variation is hidden at first, meaning that its phenotypic effects are attenuated \citep{gibson_dworkin_2004}. Hidden variation accumulates in the ancestral environment due to relaxed selection. The full, unattenuated effects of such variants can later be revealed by single mutations, by recombination into a different genetic background, or by stress-responsive developmental mechanisms \citep{gibson_dworkin_2004, hayden_etal_2011, duveau_felix_2012}.  

Hidden variation can sometimes be tracked down to cryptic genetic sequences, \textit{i.e.} to genes or portions of genes whose current effects on phenotype are attenuated relative to the magnitude of their possible effects. Attenuation may occur, for example, when a sequence is only rarely transcribed or translated \citep{rajon_masel_2011}. With such low expression levels and hence such weak selection, cryptic sequences are more prone to accumulate mutations than are sequences whose effects are not attenuated. Cryptic sequences nevertheless maintain the capacity for larger effect later, if and when attenuation is lost, \textit{e.g.} when they become constitutively expressed rather than expressed only occasionally. When cryptic sequences have mutated over long periods of time, their co-option can result in large phenotypic changes and allow for innovations that would otherwise occur on very large timescales, or might even not occur at all \citep{whitehead_etal_2008}. 

Here we explore an abstract model of cryptic sequences, their attenuation, accumulation and eventual potential co-option. Our model is inspired by the concrete example of cryptic DNA sequences that are only rarely translated into proteins. These underexpressed sequences can be whole genes or parts of genes. For example, sequences in the 3$'$ untranslated region (3$'$UTR) of a gene are only expressed when a stop codon is misread, resulting in an elongated protein \citep{rajon_masel_2011}. The frequency of elongated proteins in the cell -- and hence their overall phenotypic effect -- is normally small. Co-option occurs if a mutation changes the stop codon into a sense codon, so the 3$'$UTR becomes constitutively expressed. This has happened many times during the evolutionary history of \textit{Saccharomyces} \citep{giacomelli_etal_2007}, rodents \citep{giacomelli_etal_2007} and prokaryotes \citep{vakhrusheva_etal_2011}. Introns can similarly be co-opted when a mutation alters splicing efficiency \citep{modrek_lee_2002, kondrashov_koonin_2003, lee_etal_2012}. Similarly, constitutively down-regulated genes in regulatory networks can act as cryptic sequences. Co-option in this case can occur through mutations within a gene's regulatory region, which may lead to large increases in its expression \citep{tirosh_etal_2009, cheung_etal_2010,tirosh_etal_2010a}. 

There are other sequences that yield abundant gene products, but are nevertheless cryptic in the sense captured by our model, as a result of epistatic interactions that attenuate their phenotypic effects \citep{segre_etal_2004, hansen_2006, carlborg_etal_2006, lerouzic_carlborg_2007, phillips_2008}: co-option in this case may occur via mutations modifying epistasis. For example, mutations in two genes explain most of the evolution of light coloration in oldfield mice after the colonization of a new sandy habitat in the gulf coast of Florida \citep{steiner_etal_2007}. The interaction between a ligand (encoded by the \textit{Agouti} gene) and its receptor (\textit{Mc1r}) controls the color of the dorsal coat in mice. \textit{Mc1r} is found at the surface of pigment-producing cells and governs the relative abundance of a light pigment. In the ancestral mouse population living on a dark soil, an allele of \textit{Agouti} masks the effect of a mutation of \textit{Mc1r}, which would otherwise increase the production of the light pigment \citep{barrett_schluter_2008}. A mutation suppressing the repressing action of \textit{Agouti} increases the production of this pigment, resulting in a lighter color favored by selection on the lighter soil of the coast \citep{steiner_etal_2007}. 

\begin{figure}[!ht]
\centering
\includegraphics[width=140mm]{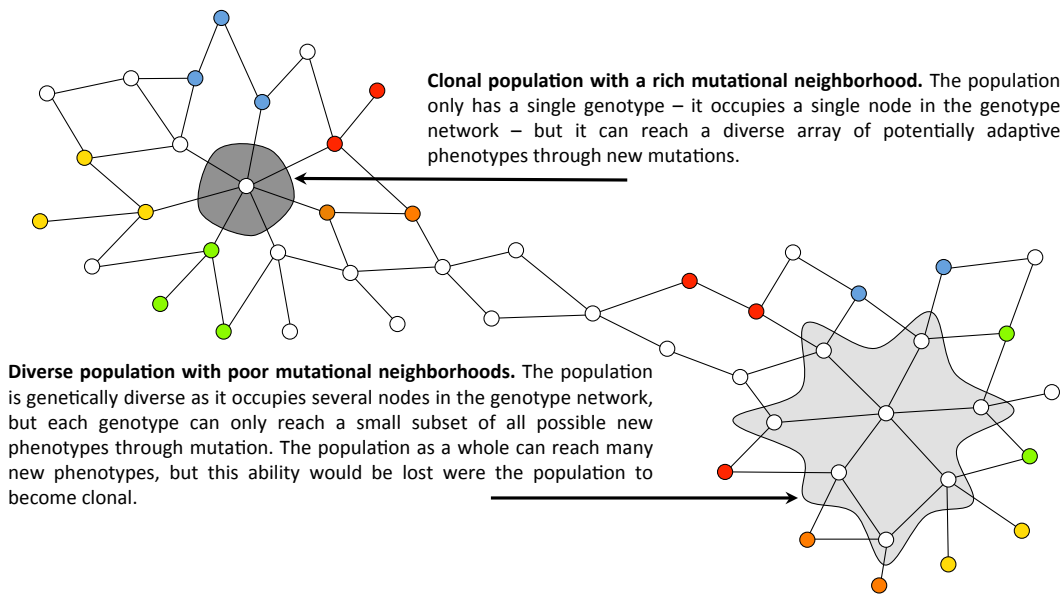}
\caption{A schematic genotype network where a clonal and a genetically diverse population can access similar sets of phenotypes through mutation. Nodes represent genotypes, which may yield different phenotypes represented by different colors (including white, which is optimal in the current environment). Connections between nodes represent possible mutations. The genotypes present in the two example populations are represented by grey areas. The population on the left is clonal, so its phenotypic potential (the distribution of new phenotypes accessible through mutations) corresponds to the neighborhood richness of the unique genotype. The population on the right is diverse, but each genotype has relatively poor neighborhood richness (genotypes can access 1-2 new phenotypes through 1 mutation). In this population, the phenotypic potential depends on neighborhood richness and on genetic diversity.}
\end{figure}

Here we investigate the adaptive potential made possible through the co-option of cryptic genetic sequences. Adaptive potential is often illustrated by genotype networks, where nodes represent genotypes and edges represent single mutational steps \citep[][see Fig. 1]{wagner_2005}. The number of new phenotypes accessible by a single mutation has two components \citep{masel_trotter_2010, wagner_2011}. First, a population that occupies many nodes on the network of possible genotypes -- \textit{i.e.} that has high genetic diversity -- may be able to access different phenotypes from each of the genotypes to which it has already spread (Fig. 1 -- right part of the network). This may increase the speed of adaptation when ``mutational neighborhoods'' are poor \citep{draghi_etal_2010, hayden_etal_2011}, that is, when each genotype can access very few new phenotypes through mutation. Second, even a clonal population -- which occupies a single node -- may have a high adaptive potential if its mutational neighborhood contains diverse phenotypes (Fig. 1 -- left part of the network). 

Genotype network models, including the special case of a ``neutral network'', are most often used to represent single proteins or single RNA sequences, where the genotype of interest can mutate into few readily accessible alternative phenotypes. These single locus genotypes generally have poor mutational neighborhoods. In contrast, quantitative phenotypes are generally affected by multiple genes \citep{visscher_2008, ehrenreich_etal_2010, ehrenreich_etal_2012, flint_mackay_2009, buckler_etal_2009}, including multiple cryptic sequences \citep{lauter_doebley_2002, gibson_dworkin_2004, rajon_masel_2011}. When quantitative traits are considered, some genotypes may have access to a great variety of potentially adaptive quantitative phenotypes through the co-option of different cryptic sequences (Fig. 2). With such a rich mutational neighborhood already available, genetic diversity might not make a very larger further contribution to the diversity of phenotypes accessible by co-option.  

Here we calculate the relative contributions of neighborhood richness and genetic diversity under a range of realistic parameter values. We find that neighborhood richness dominates for phenotypes influenced by several cryptic sequences. This is because these cryptic sequences can have substantial effect sizes, due to a history of compensatory evolution. Co-option can convert cryptic attenuated effects into large phenotypic changes and facilitate major innovations, which occur more quickly with co-option than with regular mutations to cryptic or non-cryptic sequences. Because this adaptive potential resides in genomes instead of populations, it is unaffected by losses of genetic variation, and may be particularly important when such losses are frequent.

\bigskip
\begin{SCfigure}[50][!ht]
\centering
\includegraphics[width=87mm]{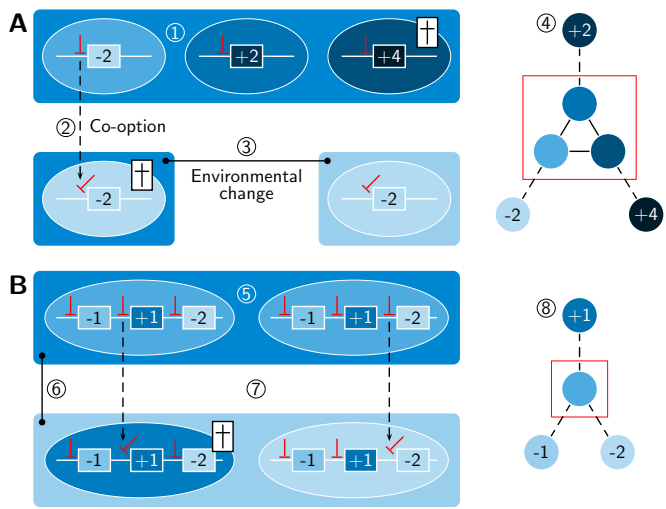}
\caption{A schematic genotype network where a clonal and a genetically diverse population can access similar sets of phenotypes through mutation. Nodes represent genotypes, which may yield different phenotypes represented by different colors (including white, which is optimal in the current environment). Connections between nodes represent possible mutations. The genotypes present in the two example populations are represented by grey areas. The population on the left is clonal, so its phenotypic potential (the distribution of new phenotypes accessible through mutations) corresponds to the neighborhood richness of the unique genotype. The population on the right is diverse, but each genotype has relatively poor neighborhood richness (genotypes can access 1-2 new phenotypes through 1 mutation). In this population, the phenotypic potential depends on neighborhood richness and on genetic diversity.}
\end{SCfigure}

\section*{Methods}

\paragraph*{Model.}

Let the phenotype, a set of $C$ quantitative characters, be determined by $L$ distinct sequences in the genome. Each sequence contributes to the $C$ characters and hence specifies a vector in a $C$-dimensional phenotype space. In its non-cryptic state, the $l^{th}$ sequence of genotype $j$ has a quantitative effect $\beta_{jlc}$ on the $c^{th}$ character. The phenotypic effect of each cryptic sequence is attenuated by a factor $\rho$, so for genotype $j$ character $c$ has a value
	\begin{equation}\label{eq:xi}
		x_{jc}= \displaystyle \sum_{l=1}^{L} \rho \times \beta_{jlc}.
	\end{equation}

Mutations occur with probability $\mu=1/(N \times 100)$ per nucleotide per generation at the $60$ nucleotides of each of the $L$ sequences, where $N$ is the population size. This assumption, which keeps constant the input of mutations per generation, is empirically supported by the decrease of $\mu$ with effective population size across species \citep{lynch_2010, sung_etal_2012}. A mutation in sequence $l$ changes the genotype $j$, and hence the quantitative effects $\beta_{jlc}$ for all $c \in [1,C]$. Most quantitative genetics models assume that mutations have a mean effect of $0$ \citep{lande_1976}. This assumption causes the probability distribution of $\beta_{jlc}$ to show an unbounded increase in variance over time \citep{lande_1976, lynch_gabriel_1983}. To avoid this unrealistic outcome, we instead assume that the mutation of sequence $l$ adds to each of the $\beta_{jlc}$ an amount sampled independently from a normal distribution with mean $-\frac{\beta_{jlc}}{50}$ and standard deviation $0.5$. This introduces a bias in the mean mutational effect, such that whenever a given sequence has a large effect size $\beta_{jlc}$, mutations will tend to decay this effect size back toward smaller values. Eventually, $\beta_{jlc}$ values reach a stationary probability distribution, whose variance can be calculated analytically for the special case where changes in $\beta_{jlc}$ are neutral \citep[dashed line at top of Fig. 3;][]{rajon_masel_2011}.

\begin{SCfigure}[50][!ht]
\centering
\includegraphics[height=87mm, angle=-90]{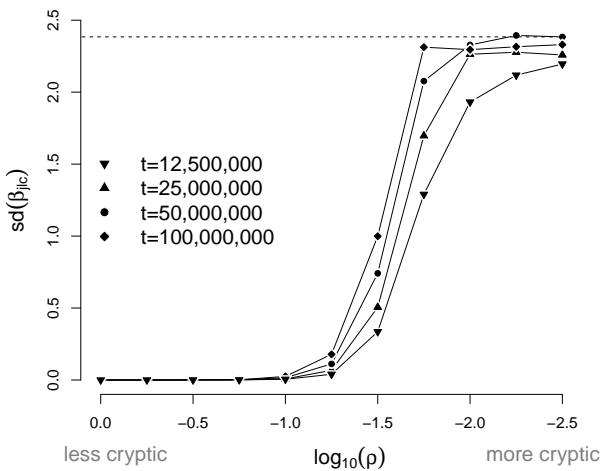}
\caption{The standard deviation of the phenotypic effects, $sd(\beta_{jlc})$, increases as $\rho$ decreases and sequences become more cryptic. Variation across sites accumulates through compensatory evolution, which occurs at a speed that depends on the strength of selection. Accordingly, when selection is strong ($\rho$ is high, crypticity is low) $sd(\beta_{jlc})$ remains low regardless of the simulation time $t$, and is unlikely to reach high values at evolutionarily relevant timescales. On the other hand, for highly cryptic sequences where selection is very weak, $sd(\beta_{jlc})$ quickly reaches high values close to the expected value in a neutral model, calculated according to eq. (S3) in \citet{rajon_masel_2011} (dashed line). Only at intermediate levels of crypticity does the variation across sites increase noticeably with evolutionary time. The average of $sd(\beta_{jlc})$ was calculated across traits in a given individual, then across individuals and across independently evolved populations.}
\end{SCfigure}

Simulations begin with all $\beta_{jlc}$ values equal to zero. Given infinite time for compensatory evolution to occur, they will asymptote to the neutral stationary probability distribution, even in the absence of crypticity. But evolutionary timescales, however long, are far from infinite. In Figure 3 we show that even over quite long evolutionary timescales, the variance in $\beta_{jlc}$ values remains far from the asymptote. We see a switch-like dependence on a key model parameter, where low levels of crypticity lead to little evolutionary change in $\beta_{jlc}$ values, while high levels of crypticity lead to $\beta_{jlc}$ values of a magnitude not far from the neutral expectation. This pattern is robust to the precise duration of the evolutionary simulations, with the cutoff level of crypticity for this transition depending only modestly on the number of generations. In subsequent figures, we set the number of generations equal to $5 \times 10^7$.

The fitness $\omega(j)$ of genotype $j$ is a product of $C$ Gaussians, each a function of character value $x_{jc}$, with optimum $0$ and variance $1$. In other words, fitness is a multivariate Gaussian with optimum $(0,...,0)$. 
For each replicate simulation in this ancestral selective environment, we simulated an asexual Wright-Fisher process over $5 \times 10^7$ generations, starting with a clonal population with all $\beta_{jlc}=0$. The population is replaced each generation by sampling, with the probability that a given new individual has genotype $j$ equal to:
	\begin{equation}
		\frac{n_j \omega(j)}{\displaystyle \sum_i ( n_i \omega(i))},
	\end{equation}			
where $n_j$ denotes the number of individuals with genotype $j$ and the sum is made over all the genotypes in the population. Each of the sampled individuals is then subjected to possible mutation: since mutational effects lie on a continuum, we have infinite alleles at each locus, and each mutation introduces a new genotype in the population. Our model maps the discrete set of genotypes present at any moment in time to a continuous $C$-dimensional phenotypic space.

Consider an adaptive challenge at the end of this simulated evolution. By co-opting a cryptic sequence, a new phenotype can be generated. The co-option of the $l^{\text{th}}$ sequence in genotype $j$ changes the value of character $x_{jc}$ to:
			\begin{equation}\label{eq:eijkl}				
				e_{jlc}= x_{jc} + (1-\rho) \times \beta_{jlc}.
			\end{equation}

The coordinates $e_{jl \bullet}=(e_{jl1}, ..., e_{jlC})$ define a point in phenotype space formed by all $C$ characters, corresponding to the phenotype that can be reached through the co-option of sequence $l$ in genotype $j$. At the end of each replicate simulation, we calculated the average Euclidean distance $d_G$ between a pair of individuals each generated via the co-option of one sequence in the same genotype $j$. We also calculated the average distance $d_P$ between two individuals of any genotype in the evolved population, again with one sequence co-opted per individual (see Appendix).
		
\paragraph*{Simulations with a new optimum.}

We measured evolvability by sampling fixation events arising from co-option and regular mutants. We used $C=3$ and generated $20$ new phenotypic optima uniformly distributed on a sphere of diameter $d$ (R script available upon request); $d$ is thus the distance to the new optimum. \\
We sampled $1000$ individuals per population at the end of the simulated evolutionary process described in the previous section. Each of these individuals generated two mutants, one with a co-option mutation and one with a regular mutation. The sequence affected by these mutations was chosen randomly among the $L$ loci. Each 
evolved population was confronted with each new optimum. For each regular or co-opted mutant, we calculated the fixation probability based on the selection coefficient, which we calculated as:
			\begin{equation}\label{eq:s}				
				s(\text{mutant})=\frac{\omega(\text{mutant})}{<\omega >}-1,
			\end{equation}
where $\omega(\text{mutant})$ is the mutant fitness and $<\omega>$ the population mean fitness. The probability of fixation for a haploid population of size $N$ is then calculated as:
			\begin{equation}\label{eq:P}				
				p_{\text{fix}}(\text{mutant})=\dfrac{1-e^{-2 s(\text{mutant})}}{1-e^{-2 N s(\text{mutant})}}.
			\end{equation}
			
To simulate the expected number of trials until fixation, we repeatedly sampled one of the $1000$ mutants from a given population, and then allowed it to fix with probability $p_\text{fix}$, until the first successful fixation occurred. The process was repeated $100$ times per evolved population for each of the 20 new optima. At the end of each repetition, we recorded the number of trials before fixation -- we report the median of this number in Figs. 7A and 8A. To reduce computation time, we used one set of mutants per evolved population, resampling from them and their fixation probabilities until the first success. Also for time efficiency, the process was stopped if no success was obtained after $10^5$ trials. This did not affect our results based on the median. For each mutant that fixed in our simulations, we also recorded the distance to the new optimum -- we report the average of this distance in Figs. 7B and 8B.

\section*{Results}

\paragraph*{Compensatory evolution creates phenotypically rich mutational neighborhoods available through co-option.}

We simulated the evolution of genotypes with single or multiple cryptic sequences contributing to a set of traits, over $50$ million generations. At the end of this simulated evolution, we quantified the phenotypic effect of co-opting one cryptic sequence. A population may have access to a variety of phenotypes through co-option. We quantified this phenotypic richness as the average Euclidean distance, in the phenotypic space formed by the $C$ characters, between two individuals subject to independent cryptic sequence co-option events (see Appendix).

In the one-locus version of our model (Fig. 4, $L=1$), the phenotypic richness accessible to the population as a whole ($d_P$) increases when the expressivity of cryptic sequences, $\rho$, decreases. This is because accumulated genetic diversity -- the variation in the effect size of a given sequence, across individuals in the population -- increases as sequences become more cryptic. With one locus, the neighborhood richness available via co-option is trivially equal to zero, since the only way to sample two mutational neighbors of the same genotype is to sample the same co-option event, at the only available locus, twice. Neighborhood richness values can be higher only when multiple different cryptic sequences are available for co-option. Accordingly, in the multi-locus model, both neighborhood richness (quantified by $d_G$) and genetic diversity contribute to the phenotypic potential of the population ($d_P$). $d_P$ includes both neighborhood richness and genetic diversity, so it is always higher than $d_G$, which captures only the former. The ratio $d_G/d_P$ quantifies their relative contributions. This ratio approaches $1$ when $\rho$ is lower than $0.1$ (Fig. 4, $L=10$). This is a very reasonable parameter range for a cryptic sequence, meaning that most of the phenotypic variability of the population resides in neighborhood richness, not genetic diversity. 

\begin{SCfigure}[50][!ht]
\centering
\includegraphics[height=87mm, angle=-90]{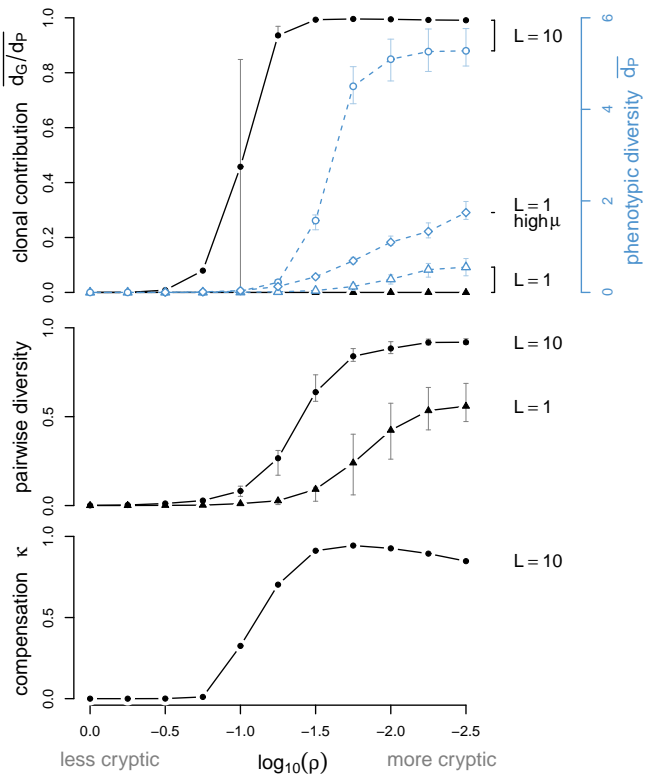}
\caption{Neighborhood richness $d_G$ explains most of the phenotypic potential of a population $d_P$ when sequences are strongly cryptic. Top: When $L=10$, both the ratio $\overline{d_G/d_P}$ and $d_P$ increase when $\rho$ decreases and sequences are more cryptic. Crypticity needs to be more complete in order to drive $d_P$ up when $L=1$, a situation in which $d_G$ always equals $0$ because an increase of neighborhood richness via intragenomic diversity is impossible. $d_P$ remains low when $L=1$, even when the effective mutation rate is increased tenfold (`high $\mu$', blue diamonds), such that the mutation rate per genome equals that for $L=10$. Middle: The large ratio $\overline{d_G/d_P}$ cannot be explained by a lack of genetic diversity in the population. Pairwise diversity was quantified as the probability that two individuals have different genotypes, calculated as $1-\sum_i f_i^2$, with $f_i$ the frequency of genotype $i$. Bottom: $\kappa$ is a metric of compensatory evolution (see text), calculated as the normalized difference between the mean variance in $\sum_l \beta_{jlc}$, expected if loci had evolved independently, and the mean variance observed in simulations. Positive values of $\kappa$ indicate compensatory evolution. Parameter values: $N=10^5$, $\mu=10^{-7}$, $C=3$. Results are averaged over $200$ $(L=1)$ or $49$ $(L=10)$ simulations. The bars in the top and center panels represent the $0.25$ and $0.75$ quantiles in the distribution of $\overline{d_G/d_P}$ across replicate simulations.}
\end{SCfigure}

The predominant role of neighborhood richness could be explained by a limited number of genotypes per population, each with high genetic potential. To examine this possibility, we quantified pairwise diversity as the probability that two individuals share the same genotype (Fig. 4). This measure of diversity increases as $\rho$ decreases, and is high when $\overline{d_G/d_P}$ is high. High values of this ratio are therefore not due to a lack of genetic polymorphism. 

In our model, the mutation rate per character increases with the number of sequences $L$.  Although this assumption seems reasonable, it might provide an artefactually lower value of $d_P$ when $L=1$. We therefore ran simulations with $L=1$ with a 10-fold higher mutation rate, for comparison to the $L=10$ case in Fig. 4, holding constant the mutation rate per genome (instead of per cryptic sequence). Pairwise diversity is superimposable for different values of $L$. With the higher mutation rate, $d_P$ increases significantly for $L=1$. Nevertheless, $d_P$ for $L=1$ remains a small fraction of the value obtained when $L=10$ (Figure 4 top panel, `high $\mu$'), so the increased adaptive potential in the multi-locus case cannot be explained by more frequent mutations alone.
 
Compare total population variability $d_P$ in the one-locus model to the ratio $d_G/d_P$ in the 10-locus case (Fig. 4). Neighborhood richness represented by the ratio builds-up even at levels of crypticity that are too low (\textit{i.e.} $\rho$ is too high) to allow substantial one-locus genetic diversity. Something is happening in the multi-locus case that is not a simple extrapolation of the one-locus case. We attribute this to compensatory evolution, whereby the fitness decrease associated with a mutation is compensated by one or several mutations at other sites \citep[Fig. 5;][]{poon_otto_2000, poon_chao_2005, rokyta_etal_2002, harcombe_etal_2009, meer_etal_2010}. When a deleterious variant is cryptic, relaxed selection increases the mean time it remains in the population. This increases the probability that a compensatory mutation will occur during its sojourn \citep{kimura_1985, phillips_1996, haag_2007}. Combinations of compensatory substitutions are therefore expected to fix more readily at low values of $\rho$, when selection on cryptic sequences is relaxed. This is consistent with observed high values of $d_G/d_P$ as well as $d_G$ when $\rho$ is small, under the influence of compensatory evolution.

\bigskip
\begin{figure}[!ht]
\centering
\includegraphics[width=110mm]{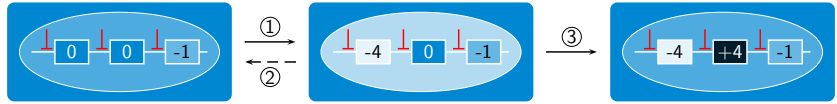}
\caption{The compensatory evolution of multilocus characters increases neighborhood richness. A mutation in a sequence controlling color (see Fig. 2) likely moves the phenotype away from the environmental optimum (1). Before this variant is eliminated, a backward mutation (2, unlikely) or a compensatory mutation (3) may occur and cancel its phenotypic effect. After the compensatory pair has fixed, a greater diversity of phenotypes can be accessed through co-option.}
\end{figure}
\medskip

Compensatory evolution means that a locus will evolve an effect size in a different direction to (i.e. negatively correlated with) the effect sizes at other loci. We can see this compensation directly, using a simple test of independence. Consider vectors of $\beta$-values, both for one locus and for an individual with $L$ loci. With independence across loci, the variance across $L$-loci genotypes is expected to be $L$ times the per-locus variance. With compensatory evolution and resulting negative correlation, the observed variance will be lower than the expectation under independence. In the bottom panel of Figure 4, we plot the metric $\kappa$, which quantifies the departure of the variance in $\sum_l \beta_{jlc}$ from its expected value under independence. Given the analytically known expectation of zero for $\sum_l \beta_{jlc}$, for each trait we calculate the observed variance as the simple mean of $(\sum_l \beta_{jlc})^2$ across all individuals in all simulations. The expected variance under independence, given that $E(\beta_{jlc})=0$, equals $L$ times the mean of $\beta_{jcl}^2$ across all simulations, individuals, and loci. $\kappa$ is the mean, across traits, of the difference between the expected variance in $\sum_l \beta_{jlc}$ and its observed value, divided by the former, such that compensatory evolution is detected by positive values of $\kappa$. In the bottom panel of Fig. 4, we see that compensatory evolution indeed occurs at low values of $\rho$, when selection on cryptic sequences is weak. However, evolution is never completely neutral (i.e. $\kappa$ is never equal to $0$) even for the most cryptic sequences considered: $\kappa$ decreases only slowly with decreasing $\rho$ when $\rho<10^{-1.75}$, and remains high even when $\rho=10^{-2.5}$. This shows that compensatory evolution is important for a wide range of cryptic sequences, not just those near the threshold at which cryptic sequences evolve away from their initial values of zero.

Selection becomes effective for values of the selection coefficient below $1/N$. For cryptic sequences, the threshold for selection is $1/(\rho N)$. As expected, $\rho$ and $N$ have similar effects (Fig. 6A). An increase in $N$, by increasing the effectiveness of selection, also decreases the sojourn time of a deleterious variant. In consequence, compensatory evolution becomes less likely when $N$ is large, and $d_G$ decreases (along with $d_G/d_P$, Fig. 6A). 

\bigskip
\begin{SCfigure}[50][!ht]
\centering
\includegraphics[height=87mm, angle=-90]{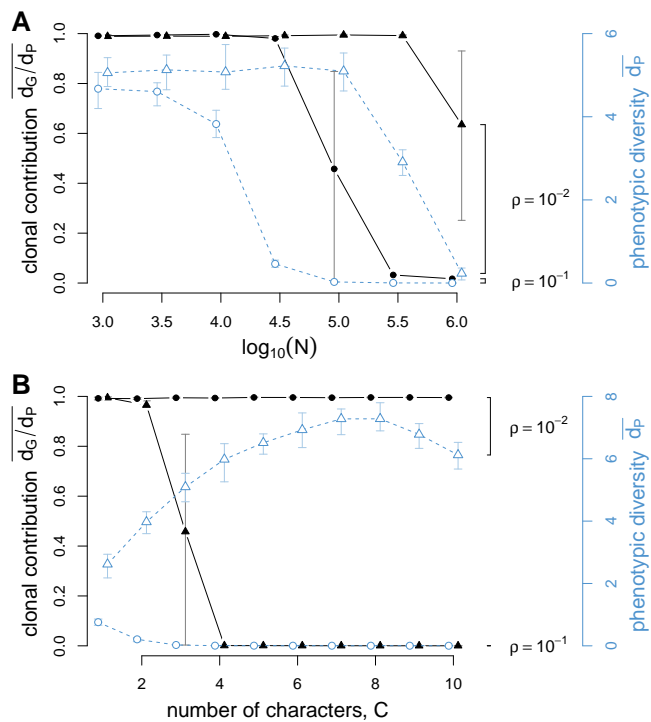}
\caption{The mean proportion of variation due to neighborhood richness ($d_G/d_P$) decreases with the population size $N$ and when weakly cryptic sequences encode a large number of characters. A: $d_G/d_P$ decreases when the product $\rho N$ exceeds a threshold.  B: $d_G/d_P$ and $d_P$ decrease with $C$ when $\rho=10^{-1}$ but not when $\rho=10^{-2}$. The blue dashed lines represent the absolute values of $\bar{d_P}$. Parameter values: $L=10$, $C=3$ (A), $N=10^5$ (B). Results are averaged over $49$ or $34$ simulations (the latter when $N > 10^5$ in panel A and $C > 5$ in panel B). The bars represent the $0.25$ and $0.75$ quantiles in the distribution of $\overline{d_G/d_P}$ across replicate simulations.}
\end{SCfigure}
\bigskip

We expect more compensatory combinations to fix when compensatory mutations are more common. When the number of characters increases, the probability that a mutation has compensating effects on all the appropriate characters becomes vanishingly small. Accordingly, the neighborhood richness $d_G$ becomes a less important part of the total variation present in the population as the number of characters increases, given substantial cryptic selection (Fig. 6B, $\rho=10^{-1}$). However, with greater crypticity ($\rho=10^{-2}$), neighborhood richness continues to dominate, even for larger numbers of character dimensions.

\paragraph*{Compensatory evolution increases evolvability.}

$d_P$ gives an easily decomposable but indirect measure of evolvability. To estimate population evolvability more directly, we assigned a new optimum phenotype at a distance $d$ (see Methods, section ``Simulations with a new optimum''). This is a $C$-dimensional version of Fisher's geometric model, where mutations correspond to vectors in $C$-dimensional phenotype space. Evolvability means the ability to generate and fix adaptive mutants. 

For each value of $\rho$, we generated a set of mutants from random individuals in each evolved population, and calculated their selection coefficients and fixation probabilities. Using Monte-Carlo simulations, we then calculated the median number of mutants trialed until one fixed (see Methods). The general trend that we observe is that as $d$ gets larger, the median number of trials approaches an asymptotic value of $2$. This is a standard asymptotic result for Fisher's geometric model \citep{fisher_1930, poon_otto_2000}: when the distance to the optimum is very large relative to the mutational effect size, nearly half the mutations will be improvements. 
Regular mutations have moderate effect sizes, so the median number of trials approaches $2$ for intermediates distances to the optimum (Fig. 7A). 

\begin{SCfigure}[50]
\centering
\includegraphics[height=87mm, angle=-90]{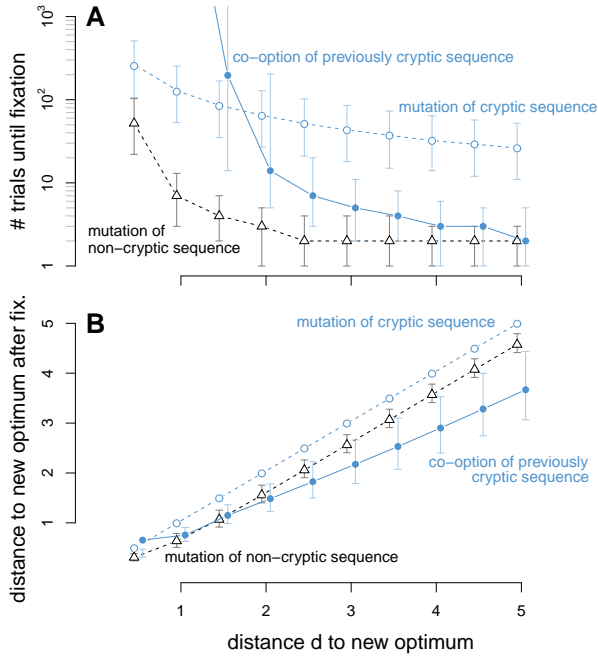}
\caption{Co-option increases the potential for adaptation to large environmental changes only. A: Waiting time for an adaptive fixation scaled according to the number of mutations trialed before one fixes -- the median number of trials is represented. A short waiting time (\textit{i.e.} a small number of trials) indicates high evolvability. Mutations in non-cryptic sequences yield a fixation event more rapidly, although the calculations do not include the fact that only co-option mutations have been pre-screened for strong, unconditionally deleterious effects  \citep{masel_2006, rajon_masel_2011}. B: Distance to the new optimum, following a fixation event. For larger environmental changes, one fixed co-option mutation yields a greater advantage than a regular mutation in either a cryptic or a non-cryptic sequence. Parameter values: $N=10^5$, $\mu=10^{-7}$, $C=3$, and $\rho=10^{-2}$ for cryptic sequences ($\rho=1$ otherwise). The bars represent the $0.25$ and $0.75$ quantiles in the distribution of the number of trials until fixation (panel A) or of the distance to the new optimum (panel B).}
\end{SCfigure}

Co-option mutations have large effect sizes. Therefore, they sometimes lead to phenotypic changes much larger than those required for small-scale adaptation -- a phenomenon called ``overshooting'' \citep{sellis_etal_2011}. Consequently, the median number of co-option trials approaches the asymptote expectation of $2$ at larger values of $d$, relative to mutations of non-attenuated effects (Fig. 7A). However, when the co-option mutations do fix, they bring the phenotype closer to the optimum than regular mutations do (Fig. 7B). 

Consider again our example of cryptic coding sequences in $3'$-UTRs. In this case, adaptation can happen in the main coding sequence ($\rho=1$) or via co-option; mutations within cryptic $3'$-UTR sequences are usually ignored as a source of adaptation. In contrast, mutations in other systems (\textit{e.g.} gene networks with complex epistasis and/or genes expressed at constitutively low levels) may frequently be cryptic to a greater or lesser extent, with no alternative non-cryptic sequence available for comparison. In Figure 7 (blue circles and dashed line) we see that even if they occur reasonably frequently, mutations within cryptic sequences are only important -- relative to co-option -- for very small changes to the optimal phenotype.

Figure 4A showed how the phenotypic potential of co-option mutations increases with crypticity, with the transition occurring at lower crypticity when compensatory evolution occurs (\textit{i.e.} with more loci). In Figure 8A we see that the parameter range in which we find high neighborhood richness corresponds to parameter range of high evolvability as assayed by our more direct measure. Compensatory evolution occurring at low $\rho$ consistently reduces the number of trials needed before an adaptive fixation. It also brings the new phenotype dramatically closer to the new optimum (Fig. 8B). In contrast, mutations of attenuated effect contribute less to evolvability (Fig. 8, dashed lines). Mutations in non-cryptic sequences are captured by the special case of $\rho=1$ (dashed line, far left).

\begin{SCfigure}[50][!ht]
\centering
\includegraphics[height=87mm, angle=-90]{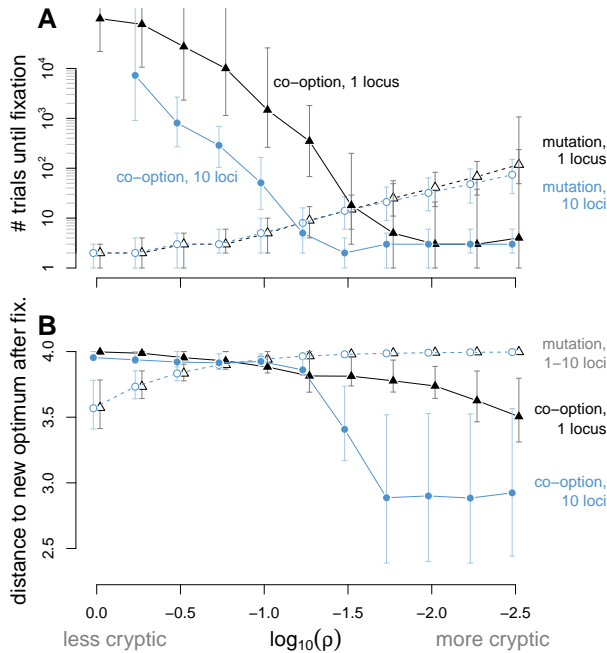}
\caption{Higher crypticity leads to higher evolvability through co-option. A: The evolvability of co-option mutations, measured as the median of the (small) number of mutants that need to be trialed before one fixes, mirrors the results of Fig. 4A. This means that high evolvability tracks high neighborhood richness driven by compensatory evolution in cryptic sequences. Regular mutations in cryptic sequences are shown for comparison. B: Not only do co-option mutants fix more readily when they reveal more cryptic sequences, they also reach phenotypes much closer to the new optimum. Large adaptive phenotypic changes are possible through single co-option mutations in the $10$-loci case when sufficient variation across cryptic sequences has accumulated through compensatory evolution (\textit{i.e.} when $d_P$ and $d_G/d_P$ are high in Fig. 4A). Same parameter values as in Fig. 4; the distance from the old to the new optimum $d$ equals $4$. The bars represent the $0.25$ and $0.75$ quantiles in the distribution of the number of trials until fixation (panel A) and of the distance to the new optimum (panel B).}
\end{SCfigure}

Note that if we wanted to infer the extent to which cryptic rather than non-cryptic mutations contribute to evolvability, we would also need to know the ratio of mutation rates between co-option and ``normal'' mutations, which will vary between different biological systems. We have calculated the phenotypic potential ($d_P$) and evolvability for one value of crypticity ($\rho$) at a time, corresponding to one category of sites. 

The primary result of this paper is to infer that if one accepts that co-option mutations might be important to evolvability, then their importance is almost independent of the presence of pre-existing genetic diversity. This result overturns, for polygenic traits, the previously dominant metaphor of evolvability via the “spread” of a population across a neutral network \citep{wagner_2005, wagner_2008, draghi_etal_2010}. We attribute this result to compensatory evolution on cryptic sequences while they are cryptic, which leads to a larger eventual effect size if and when they are eventually co-opted. This prior compensatory evolution yields co-option mutations with larger effect sizes than ``normal'' mutations, with a larger resulting contribution to evolvability.

\section*{Discussion}

Cryptic sequences may exist within a range of genetic architectures, which we illustrate with two representative examples. In one example, a gene can epistatically control the effect of other genes, \textit{e.g.} through the regulation of their expression. Each gene in the epistatic network is a potentially cryptic sequence, and co-option in this system occurs via a change in epistatic interactions. In our second example ($3'$-UTRs), cryptic sequences have additive effects, and are associated with nearby non-cryptic sequences that also contribute additively to the same set of phenotypic characters. Co-option converts a cryptic sequence into a non-cryptic one. 

In both our examples, the co-option of a cryptic sequence results in larger phenotypic changes when compared to other classes of mutations. The large size of the change is a consequence of compensatory evolution that took place prior to co-option, while the sequence in question was cryptic. The possibility of selection prior to mutation has been called the “look-ahead effect” \citep{whitehead_etal_2008}, and has been previously studied in sexual \citep{masel_2006, kim_2007} as well as asexual \citep{whitehead_etal_2008, rajon_masel_2011} populations.

Many metrics of evolvability measure how frequently potentially adaptive mutants are generated, but neglect their magnitude. By allowing access to distant phenotypes in a small number of mutational steps, co-option may be an important mechanism for evolutionary innovations that involve large-scale phenotypic changes. This evolvability benefit of the co-option of cryptic sequences, which we show in Figure 8B, is in addition to other previously reported advantages (not considered in our model), in particular the prior exclusion of strongly deleterious alleles whose expression is unconditionally disadvantageous in all environments \citep{masel_2006, rajon_masel_2011}, and the ability to overcome certain forms of synergistic \citep{griswold_masel_2009} and antagonistic \citep{masel_2006, kim_2007, whitehead_etal_2008} epistasis.

While our model is reasonably general, it makes important assumptions, upon which our conclusions might depend. Some of these assumptions are mathematical conveniences not likely to alter outcomes. More significantly, we assume that i) reproduction is asexual, ii) \textit{de novo} mutations (including mutations that co-opt previously cryptic sequences), rather than standing genetic variation alone, sometimes contribute to evolvability, iii) that phenotypic traits are affected by more than one cryptic sequence, and iv) that co-option mutations in a single gene affect a relatively small number of phenotypic traits. Assumption iii) seems unproblematic in the light of numerous QTL and GWAS studies detecting large numbers of loci contributing to individual traits \citep{visscher_2008, ehrenreich_etal_2010, ehrenreich_etal_2012, flint_mackay_2009, buckler_etal_2009}. Assumption iv) is also compatible with recent findings indicating that the number of traits encoded by a given gene (pleiotropy) is restricted to a small fraction of the traits measured in yeast, nematode, fish, mice and human \citep{goh_etal_2007, wagner_etal_2008, kenney_etal_2008, albert_etal_2008, wang_etal_2010}, and organized into modules where genes contribute to similar sets of traits \citep{wang_etal_2010}. Our model describes the evolution of such a module.

In our model, adaptation proceeds via the co-option of a single cryptic sequence. This assumption seems reasonable, but note that other ``evolutionary capacitance'' systems exist where several cryptic sequences may be co-opted at once -- \textit{e.g.} the [PSI+] prion in yeast \citep{griswold_masel_2009, Torabi_kruglyak_2012} or the Rho Terminator in \textit{Escherichia coli} \citep{freddolino_etal_2012}. In such cases, the contribution of cryptic sequences to adaptation can be more extreme than the situation described here. 

Below we discuss the first two key assumptions -- asexual reproduction and adaptation from \textit{de novo} mutations -- in more detail. First, consider our assumption of asexual reproduction. Our results rely strongly on compensatory evolution, and empirical data \citep{meer_etal_2010}, as well as the simulation results presented here, suggest that compensatory evolution is common in the absence of recombination. The key issue is whether high rates of compensatory evolution are also expected among cryptic sequences in sexual populations. Unfortunately, comparable sexual simulations would require the tracking of a far larger number of genotypes, making such simulations computationally inaccessible. However, some heuristic predictions can be made, based on previous analytical theory on this topic. Consider the simple example of a segregating pair of loci with mutually compensatory phenotypic effects. When recombination breaks up compensatory pairings, alleles from rare pairs are likely eliminated. This will initially select against new compensatory allele pairs (when their component alleles are rare) but may later favor them if they survive to become common. On the other hand, two alleles that would form a compensatory pair but appeared in different individuals may sometimes be brought together by recombination. The net result of these effects is complex, but theory shows that in the slightly different case where the pair of compensatory alleles is more fit than the original genotype, pairs fix more frequently with low recombination, and equally frequently with either high or zero recombination, so long as selection against cryptic alleles is weak \citep{weissman_etal_2010}. This theoretical finding suggests that our results may apply to sexual populations as well. Indeed, empirically, compensatory evolution seems to occur despite frequent recombination in sexual species, as suggested by QTL mapping studies that have reported alleles with opposite phenotypic effects at different loci \citep{rieseberg_etal_1999, brem_kruglyak_2005, carlborg_etal_2006, visscher_2008}. 

Rare recombination can even facilitate the co-option of cryptic sequences via a different mechanism. With facultative sex, recombination breaks up compensatory combinations and can result in new phenotypes \citep{lynch_gabriel_1983}. In this case, co-option occurs via recombination whereas in our model co-option occurs through mutation \citep{masel_trotter_2010}. Co-option via recombination requires genetic diversity, whereas co-option by mutation as treated here does not. In other words, here we have shown how cryptic sequences contribute not only to standing genetic variation, but also to the effects of \textit{de novo} mutations.

This brings us to our second major assumption, namely that \textit{de novo} mutations are sometimes important to adaptation. There is support for this in natural populations of garter snakes \citep{Feldman_etal_2009}, mice \citep{linnen_etal_2009}, Drosophila \citep{karasov_etal_2010} and humans \citep{peter_etal_2012}, as well as experimental populations of maize \citep{durand_etal_2010} and bacteria \citep{lenski_travisano_1994, cooper_lenski_2010, blount_etal_2012}. Note that we do not assume that \textit{de novo} mutations are more important than standing genetic variation, merely that they are important in some instances.

There are many ways that genetic diversity can be lost. Genetic diversity is eroded by genetic drift in small populations \citep{willi_etal_2006}. In addition, populations of any size can also suffer from losses of genetic diversity when an adaptive allele sweeps to fixation and brings linked loci with it \citep{gillespie_2000, gillespie_2001}. Genetic variance then needs time to recover before it can be used again for adaptation \citep{lerouzic_carlborg_2007}. Inbreeding can increase background selection against recessive alleles and also eliminate variation at linked loci. When these stochastic processes are stronger than selection against cryptic sequences, as captured by small $N$ in Fig. 6A, neighborhood richness dominates evolutionary potential. 

It was previously thought that the diversity of phenotypes produced from cryptic variation is lost when genetic variance is lost. Without the ability to generate phenotypically diverse variants, populations facing new environmental conditions may fail to adapt. Here we have shown that there is still hope: quantitative characters are encoded by multiple sequences, each of which can have a different phenotypic effect through co-option and facilitate adaptation.

\section*{Acknowledgments}
We thank A. Le Rouzic, A. Stewart, F. Kondrashov and M. Uyenoyama for helpful comments on the manuscript, and T. Secomb for helpful discussions. J. M. is a Pew Scholar in the Biomedical Sciences and was also supported in part by a fellowship at the Wissenschaftskolleg zu Berlin.

\newpage
\section*{Appendix}
	
\paragraph*{Calculation of $d_G$ and $d_P$.}

The Euclidean distance between two points in phenotype space, corresponding to genotypes $j_1$ and $j_2$ with sequences $l_1$ and $l_2$ co-opted, equals
			\begin{equation}\label{eq:dist}				
				d_{j_1 l_1 \rightarrow j_2 l_2}=\displaystyle \sqrt{\sum_{c=1}^{C} \biggl( e_{j_1 l_1 c} - e_{j_2 l_2 c} \biggr)^2}.
			\end{equation}

The potential phenotypic range that a given genotype $j$ can access by co-option can be represented by the set of $L$ points $e_{jl \bullet}$ in the phenotype space, where each point corresponds to the co-option of one cryptic sequence. The ability of a genotype to reach new phenotypes by co-option can be summarized by the mean distance between two of these points. An individual with genotype $j$ and sequence $l_1$ co-opted is at an average distance from another individual with the same genotype and any sequence co-opted:
			\begin{equation}\label{eq:dG1}				
				\displaystyle\dfrac{1}{L} \sum_{l_2=1}^{L} \biggl( d_{j l_1 \rightarrow j l_2} \biggr).
			\end{equation}
Therefore, on average, two individuals with the same genotype (assuming they exist in the population), each with one sequence co-opted, will be at a distance:
			\begin{equation}\label{eq:dG2}				
				\displaystyle\dfrac{1}{L^2} \sum_{l_1=1}^{L} \sum_{l_2=1}^{L} \biggl( d_{j l_1 \rightarrow j l_2} \biggr).
			\end{equation}
Note that each distance between two individuals with initial genotype $j$ is averaged over $L^2$ possible values, which include $0$ when the same sequence is co-opted in two different individuals. 
At the end of a given simulation, $n_{gen}$ different genotypes segregate in the population, each in $n_{j}$ copies. Consider sampling two individuals at random in order to assess the expected pairwise differences between them following co-option, with and without the condition that the two individuals start with the same genotype. An individual with genotype $j$ may be compared to $n_j-1$ individuals that share the same initial genotype, so the total number of pairwise distances per genotype equals $n_j (n_j-1)/2$. In the whole population, the average distance between two individuals with the same initial genotype and one co-option mutation therefore equals:
			\begin{equation}\label{eq:dG2}		
				d_{G} =\displaystyle\dfrac{1}{ L^2  \times \sum_{j=1}^{n_{gen}} ( n_j (n_j-1) )} \times \sum_{j=1}^{n_{gen}} \biggl[ n_j (n_j-1) \sum_{l_1=1}^{L} \sum_{l_2=1}^{L} \biggl( d_{j l_1 \rightarrow j l_2} \biggr) \biggr].
			\end{equation}

We want to compare this distance $d_G$, which represents the potential for phenotypic evolution of any one representative genotype in the population, to the comparable distance $d_P$. An individual in the population with genotype $j_1$ and with sequence $l_1$ co-opted can be compared to $N-1$ other individuals, among which $n_{j_1}-1$ have the same initial genotype $j_1$.  Its average phenotypic distance to any other individual in the population with genotype $j_2$ and with a sequence $l_2$ co-opted thus equals:
			\begin{equation}\label{eq:dP1}
			 \displaystyle\dfrac{1}{L \times (N-1) } \biggl[ \sum_{j_2=1,j_2\neq j_1}^{n_{gen}} \biggl( n_{j_2} \sum_{l_2=1}^{L} \biggl( d_{j_1 l_1 \rightarrow j_2 l_2} \biggr) \biggr) + (n_{j_1}-1) \sum_{l_2=1}^{L} \biggl( d_{j_1 l_1 \rightarrow j_1 l_2} \biggr) \biggr],
			\end{equation}
Averaged across all sequences that could be co-opted in the first individual, the mean distance between an individual with initial genotype $j_1$ and any other individual in the population, after the co-option of one cryptic sequence in each genotype, equals:
			\begin{equation}\label{eq:dP2}
				\dfrac{1}{ L^2 \times (N-1)} \displaystyle \sum_{l_1=1}^{L} \biggl[ \sum_{j_2=1,j_2\neq j_1}^{n_{gen}} \biggl( n_{j_2} \sum_{l_2=1}^{L} \biggl( d_{j_1 l_1 \rightarrow j_2 l_2} \biggr) \biggr) 
				+ (n_{j_1}-1) \sum_{l_2=1}^{L} \biggl( d_{j_1 l_1 \rightarrow j_1 l_2} \biggr) \biggr]
			\end{equation}
Averaged over all individuals in the population, the mean phenotypic distance between any two mutated individuals in the population equals:
			\begin{equation}\label{eq:dP2}
				d_{P} =\dfrac{1}{ L^2 \times N (N-1)} \displaystyle \sum_{j_1=1}^{n_{gen}} n_{j_1} \sum_{l_1=1}^{L} \biggl[ \sum_{j_2=1,j_2\neq j_1}^{n_{gen}} \biggl( n_{j_2} \sum_{l_2=1}^{L} \biggl( d_{j_1 l_1 \rightarrow j_2 l_2} \biggr) \biggr) 
				+ (n_{j_1}-1) \sum_{l_2=1}^{L} \biggl( d_{j_1 l_1 \rightarrow j_1 l_2} \biggr) \biggr]
			\end{equation}


\newpage
\begin{thebibliography}{69}
\providecommand{\natexlab}[1]{#1}

\bibitem[{Albert et~al.(2008)Albert, Sawaya, Vines, Knecht, Miller, Summers,
  Balabhadra, Kingsley, and Schluter}]{albert_etal_2008}
Albert, A. Y.~K., S.~Sawaya, T.~H. Vines, A.~K. Knecht, C.~T. Miller, B.~R.
  Summers, S.~Balabhadra, D.~M. Kingsley, and D.~Schluter. 2008.
\newblock The genetics of adaptive shape shift in stickleback: pleiotropy and
  effect size.
\newblock Evolution 62:76--85.

\bibitem[{Barrett and Schluter(2008)}]{barrett_schluter_2008}
Barrett, R.~D., and D.~Schluter. 2008.
\newblock Adaptation from standing genetic variation.
\newblock Trends Ecol. Evol. 23:38--44.

\bibitem[{Blount et~al.(2012)Blount, Barrick, Davidson, and
  Lenski}]{blount_etal_2012}
Blount, Z.~D., J.~E. Barrick, C.~J. Davidson, and R.~E. Lenski. 2012.
\newblock Genomic analysis of a key innovation in an experimental
  \textit{Escherichia coli} population.
\newblock Nature 489:513--8.

\bibitem[{Brem and Kruglyak(2005)}]{brem_kruglyak_2005}
Brem, R.~B., and L.~Kruglyak. 2005.
\newblock The landscape of genetic complexity across 5,700 gene expression
  traits in yeast.
\newblock Proc. Natl. Acad. Sci. U. S. A. 102:1572--1577.

\bibitem[{Buckler et~al.(2009)Buckler, Holland, Bradbury, Acharya, Brown,
  Browne, Ersoz, {Flint-Garcia}, Garcia, and Glaubitz}]{buckler_etal_2009}
Buckler, E.~S., J.~B. Holland, P.~J. Bradbury, C.~B. Acharya, P.~J. Brown,
  C.~Browne, E.~Ersoz, S.~{Flint-Garcia}, A.~Garcia, and J.~C. Glaubitz. 2009.
\newblock The genetic architecture of maize flowering time.
\newblock Science 325:714.

\bibitem[{Carlborg et~al.(2006)Carlborg, Jacobsson, {A}hgren, Siegel, and
  Andersson}]{carlborg_etal_2006}
Carlborg, O., L.~Jacobsson, P.~{A}hgren, P.~Siegel, and L.~Andersson. 2006.
\newblock Epistasis and the release of genetic variation during long-term
  selection.
\newblock Nat. Genet. 38:418--420.

\bibitem[{Cheung et~al.(2010)Cheung, Nayak, Wang, Elwyn, Cousins, Morley, and
  Spielman}]{cheung_etal_2010}
Cheung, V.~G., R.~R. Nayak, I.~X. Wang, S.~Elwyn, S.~M. Cousins, M.~Morley, and
  R.~S. Spielman. 2010.
\newblock Polymorphic cis- and trans-regulation of human gene expression.
\newblock PLoS Biol 8.

\bibitem[{Cooper and Lenski(2010)}]{cooper_lenski_2010}
Cooper, T.~F., and R.~E. Lenski. 2010.
\newblock Experimental evolution with e. coli in diverse resource environments.
  i. fluctuating environments promote divergence of replicate populations.
\newblock {BMC} Evol. Biol. 10.

\bibitem[{Draghi et~al.(2010)Draghi, Parsons, Wagner, and
  Plotkin}]{draghi_etal_2010}
Draghi, J., T.~Parsons, G.~Wagner, and J.~Plotkin. 2010.
\newblock Mutational robustness can facilitate adaptation.
\newblock Nature 463:353--355.

\bibitem[{Durand et~al.(2012)Durand, Tenaillon, Ridel, Coubriche, Jamin,
  Jouanne, Ressayer, Charcosset, and Dillmann}]{durand_etal_2010}
Durand, E., M.~I. Tenaillon, C.~Ridel, D.~Coubriche, P.~Jamin, S.~Jouanne,
  A.~Ressayer, A.~Charcosset, and C.~Dillmann. 2012.
\newblock Standing variation and new mutations both contribute to a fast
  response to selection for flowering time in maize inbreds.
\newblock {BMC} Evol. Biol. 10.

\bibitem[{Duveau and F{\'e}lix(2012)}]{duveau_felix_2012}
Duveau, F., and M.-A. F{\'e}lix. 2012.
\newblock Role of pleiotropy in the evolution of a cryptic developmental
  variation in \textit{Caenorhabditis elegans}.
\newblock PLoS Biol 10:e1001230.

\bibitem[{Ehrenreich et~al.(2012)Ehrenreich, Bloom, Torabi, Wang, Jia, and
  Kruglyak}]{ehrenreich_etal_2012}
Ehrenreich, I.~M., J.~Bloom, N.~Torabi, X.~Wang, Y.~Jia, and L.~Kruglyak. 2012.
\newblock Genetic architecture of highly complex chemical resistance traits
  across four yeast strains.
\newblock {PLoS} Genet 8:e1002570.

\bibitem[{Ehrenreich et~al.(2010)Ehrenreich, Torabi, Jia, Kent, Martis,
  Shapiro, Gresham, Caudy, and Kruglyak}]{ehrenreich_etal_2010}
Ehrenreich, I.~M., N.~Torabi, Y.~Jia, J.~Kent, S.~Martis, J.~A. Shapiro,
  D.~Gresham, A.~A. Caudy, and L.~Kruglyak. 2010.
\newblock Dissection of genetically complex traits with extremely large pools
  of yeast segregants.
\newblock Nature 464:1039--1042.

\bibitem[{Feldman et~al.(2009)Feldman, {Brodie Jr.}, {Broodie III}, and
  Pfrender}]{Feldman_etal_2009}
Feldman, C.~R., E.~D. {Brodie Jr.}, E.~D. {Broodie III}, and M.~E. Pfrender.
  2009.
\newblock The evolutionary origins of beneficial alleles during the repeated
  adaptation of garter snakes to deadly prey.
\newblock Proc. Natl. Acad. Sci. USA 106:13415--13420.

\bibitem[{Fisher(1930)}]{fisher_1930}
Fisher, R.~A. 1930.
\newblock Genetical theory of natural selection.
\newblock The Clarendon Press, Oxford, U. K.

\bibitem[{Flint and Mackay(2009)}]{flint_mackay_2009}
Flint, J., and T.~F.~C. Mackay. 2009.
\newblock Genetic architecture of quantitative traits in mice, flies, and
  humans.
\newblock Genome Res. 19:723.

\bibitem[{Freddolino et~al.(2012)Freddolino, Goodarzi, and
  Tazavoie}]{freddolino_etal_2012}
Freddolino, P.~L., H.~Goodarzi, and S.~Tazavoie. 2012.
\newblock Fitness landscape transformation through a single amino acid change
  in the {Rho} terminator.
\newblock {PLoS} Genet 8:e1002744.

\bibitem[{Giacomelli et~al.(2007)Giacomelli, Hancock, and
  Masel}]{giacomelli_etal_2007}
Giacomelli, M., A.~Hancock, and J.~Masel. 2007.
\newblock The conversion of 3$'$-{UTR}s into coding regions.
\newblock Mol. Biol. Evol. 24:457--464.

\bibitem[{Gibson and Dworkin(2004)}]{gibson_dworkin_2004}
Gibson, G., and I.~Dworkin. 2004.
\newblock Uncovering cryptic genetic variation.
\newblock Nat. Rev. Genet. 5:681--690.

\bibitem[{Gillespie(2000)}]{gillespie_2000}
Gillespie, J. 2000.
\newblock The neutral theory in an infinite population.
\newblock Gene 261:11--18.

\bibitem[{Gillespie(2001)}]{gillespie_2001}
---{}---{}---. 2001.
\newblock Is the population size of a species relevant to its evolution?
\newblock Evolution 55:2161--2169.

\bibitem[{Goh et~al.(2007)Goh, Cusick, Valle, Childs, Vidal, and
  Barab{\'a}si}]{goh_etal_2007}
Goh, K.~I., M.~E. Cusick, D.~Valle, B.~Childs, M.~Vidal, and A.-L.
  Barab{\'a}si. 2007.
\newblock The human disease network.
\newblock Proc. Natl. Acad. Sci. U. S. A. 104:8685--8690.

\bibitem[{Griswold and Masel(2009)}]{griswold_masel_2009}
Griswold, C., and J.~Masel. 2009.
\newblock Complex adaptations can drive the evolution of the capacitor
  {[PSI+]}, even with realistic rates of yeast sex.
\newblock PLoS Genet. 5:e1000517.

\bibitem[{Haag(2007)}]{haag_2007}
Haag, E.~S. 2007.
\newblock Compensatory vs. pseudocompensatory evolution in molecular and
  developmental interactions.
\newblock Genetica 129:45--55.

\bibitem[{Hansen(2006)}]{hansen_2006}
Hansen, T.~F. 2006.
\newblock The evolution of genetic architecture.
\newblock Annu. Rev. Ecol. Evol. Syst. 37:123--157.

\bibitem[{Harcombe et~al.(2009)Harcombe, Springman, and
  Bull}]{harcombe_etal_2009}
Harcombe, W.~R., R.~Springman, and J.~J. Bull. 2009.
\newblock Compensatory evolution for a gene deletion is not limited to its
  immediate functional network.
\newblock {BMC} Evol. Biol. 9:106.

\bibitem[{Hayden et~al.(2011)Hayden, Ferrada, and Wagner}]{hayden_etal_2011}
Hayden, E.~J., E.~Ferrada, and A.~Wagner. 2011.
\newblock Cryptic genetic variation promotes rapid evolutionary adaptation in
  an {RNA} enzyme.
\newblock Nature 474:92--5.

\bibitem[{Karasov et~al.(2010)Karasov, Messer, and Petrov}]{karasov_etal_2010}
Karasov, T., P.~W. Messer, and D.~A. Petrov. 2010.
\newblock Evidence that adaptation in \textit{Drosophila} is not limited by
  mutation at single sites.
\newblock PLoS Genet 6:e1000924.

\bibitem[{Kenney-Hunt et~al.(2008)Kenney-Hunt, Wang, Norgard, Fawcett, Falk,
  Pletscher, Jarvis, Roseman, Wolf, and Cheverud}]{kenney_etal_2008}
Kenney-Hunt, J.~P., B.~Wang, E.~A. Norgard, G.~Fawcett, D.~Falk, L.~S.
  Pletscher, J.~P. Jarvis, C.~Roseman, J.~Wolf, and J.~M. Cheverud. 2008.
\newblock Pleiotropic patterns of quantitative trait loci for 70 murine
  skeletal traits.
\newblock Genetics 178:2275--88.

\bibitem[{Kim(2007)}]{kim_2007}
Kim, Y. 2007.
\newblock Rate of adaptive peak shifts with partial genetic robustness.
\newblock Evolution 61:1847--1856.

\bibitem[{Kimura(1985)}]{kimura_1985}
Kimura, M. 1985.
\newblock The role of compensatory neutral mutations in molecular evolution.
\newblock J. Genet. 64:7--19.

\bibitem[{Kondrashov and Koonin(2003)}]{kondrashov_koonin_2003}
Kondrashov, F.~A., and E.~V. Koonin. 2003.
\newblock Evolution of alternative splicing: deletions, insertions and origin
  of functional parts of proteins from intron sequences.
\newblock Trends Genet. 19:115--119.

\bibitem[{Lande(1976)}]{lande_1976}
Lande, R. 1976.
\newblock The maintenance of genetic variability by mutation in a polygenic
  character with linked loci.
\newblock Genet. Res. 26:221--235.

\bibitem[{Lauter and Doebley(2002)}]{lauter_doebley_2002}
Lauter, N., and J.~Doebley. 2002.
\newblock Genetic variation for phenotypically invariant traits detected in
  teosinte: Implications for the evolution of novel forms.
\newblock Genetics 160:333--342.

\bibitem[{Le~Rouzic and Carlborg(2007)}]{lerouzic_carlborg_2007}
Le~Rouzic, A., and {\"O}.~Carlborg. 2007.
\newblock Evolutionary potential of hidden genetic variation.
\newblock Trends Ecol. Evol. 23:33--37.

\bibitem[{Lee et~al.(2012)Lee, Gamazon, Rebman, Lee, Lee, Dolan, Cox, and
  Lussier}]{lee_etal_2012}
Lee, Y., E.~R. Gamazon, E.~Rebman, Y.~Lee, S.~Lee, M.~E. Dolan, N.~J. Cox, and
  Y.~A. Lussier. 2012.
\newblock Variants affecting exon skipping contribute to complex traits.
\newblock {PLoS} Genet 8:e1002998.

\bibitem[{Lenski and Travisano(1994)}]{lenski_travisano_1994}
Lenski, R.~E., and M.~Travisano. 1994.
\newblock Dynamics of adaptation and diversification: A 10,000-generation
  experiment with bacterial populations.
\newblock Proc Natl Acad Sci U S A 91:6808--6814.

\bibitem[{Linnen et~al.(2009)Linnen, Kingsley, Jensen, and
  Hoekstra}]{linnen_etal_2009}
Linnen, C.~R., E.~P. Kingsley, J.~D. Jensen, and H.~E. Hoekstra. 2009.
\newblock On the origin and spread of an adaptive allele in deer mice.
\newblock Science 325:1095--1098.

\bibitem[{Lynch(2010)}]{lynch_2010}
Lynch, M. 2010.
\newblock Evolution of the mutation rate.
\newblock Trends Genet. 26:345--352.

\bibitem[{Lynch and Gabriel(1983)}]{lynch_gabriel_1983}
Lynch, M., and W.~Gabriel. 1983.
\newblock Phenotypic evolution and parthenogenesis.
\newblock Am. Nat. 122:745--764.

\bibitem[{Masel(2006)}]{masel_2006}
Masel, J. 2006.
\newblock Cryptic genetic variation is enriched for potential adaptations.
\newblock Genetics 172:1985--1991.

\bibitem[{Masel and Trotter(2010)}]{masel_trotter_2010}
Masel, J., and M.~V. Trotter. 2010.
\newblock Robustness and evolvability.
\newblock Trends in Genetics 26:406--414.

\bibitem[{Meer et~al.(2010)Meer, Kondrashov, {Artzy-Randrup}, and
  Kondrashov}]{meer_etal_2010}
Meer, M.~V., A.~S. Kondrashov, Y.~{Artzy-Randrup}, and F.~A. Kondrashov. 2010.
\newblock Compensatory evolution in mitochondrial t{RNA}s navigates valleys of
  low fitness.
\newblock Nature 464:279--283.

\bibitem[{Modrek and Lee(2002)}]{modrek_lee_2002}
Modrek, B., and C.~Lee. 2002.
\newblock A genomic view of alternative splicing.
\newblock Nat. Genet. 30:13--19.

\bibitem[{Peter et~al.(2012)Peter, Huerta-Sanchez, and
  Nielsen}]{peter_etal_2012}
Peter, B.~M., E.~Huerta-Sanchez, and R.~Nielsen. 2012.
\newblock Distinguishing between selective sweeps from standing variation and
  from a \textit{de novo} mutation.
\newblock PLoS Genet 8:e1003011.

\bibitem[{Phillips(1996)}]{phillips_1996}
Phillips, P.~C. 1996.
\newblock Waiting for a compensatory mutation: phase zero of the
  shifting-balance process.
\newblock Genet. Res. 67:271--283.

\bibitem[{Phillips(2008)}]{phillips_2008}
---{}---{}---. 2008.
\newblock Epistasis - the essential role of gene interactions in the structure
  and evolution of genetic systems.
\newblock Nat. Rev. Genet. 9:855--867.

\bibitem[{Poon and Chao(2005)}]{poon_chao_2005}
Poon, A., and L.~Chao. 2005.
\newblock The rate of compensatory mutation in the {DNA} bacteriophage
  {$\phi$X174}.
\newblock Genetics 170:989.

\bibitem[{Poon and Otto(2000)}]{poon_otto_2000}
Poon, A., and S.~Otto. 2000.
\newblock Compensating for our load of mutations: freezing the meltdown of
  small populations.
\newblock Evolution 54:1467--1479.

\bibitem[{Rajon and Masel(2011)}]{rajon_masel_2011}
Rajon, E., and J.~Masel. 2011.
\newblock Evolution of molecular error rates and the consequences for
  evolvability.
\newblock Proc. Natl. Acad. Sci. USA 108:1082--1087.

\bibitem[{Rieseberg et~al.(1999)Rieseberg, Archer, and
  Wayne}]{rieseberg_etal_1999}
Rieseberg, L.~H., M.~A. Archer, and R.~K. Wayne. 1999.
\newblock Transgressive segregation, adaptation and speciation.
\newblock Heredity 83:363--372.

\bibitem[{Rokyta et~al.(2002)Rokyta, Badgett, Molineux, and
  Bull}]{rokyta_etal_2002}
Rokyta, D., M.~R. Badgett, I.~J. Molineux, and J.~J. Bull. 2002.
\newblock Experimental genomic evolution: extensive compensation for loss of
  {DNA} ligase activity in a virus.
\newblock Mol. Biol. Evol. 19:230.

\bibitem[{Segre et~al.(2004)Segre, {DeLuna}, Church, and
  Kishony}]{segre_etal_2004}
Segre, D., A.~{DeLuna}, G.~M. Church, and R.~Kishony. 2004.
\newblock Modular epistasis in yeast metabolism.
\newblock Nat. Genet. 37:77--83.

\bibitem[{Sellis et~al.(2011)Sellis, Callahan, Petrov, and
  Messer}]{sellis_etal_2011}
Sellis, D., B.~J. Callahan, D.~A. Petrov, and P.~W. Messer. 2011.
\newblock Heterozygote advantage as a natural consequence of adaptation in
  diploids.
\newblock Proc Natl Acad Sci U S A 108:20666--71.

\bibitem[{Steiner et~al.(2007)Steiner, Weber, and Hoekstra}]{steiner_etal_2007}
Steiner, C.~C., J.~N. Weber, and H.~E. Hoekstra. 2007.
\newblock Adaptive variation in beach mice produced by two interacting
  pigmentation genes.
\newblock {PLoS} Biol. 5:e219.

\bibitem[{Sung et~al.(2012)Sung, Ackerman, Miller, Doak, and
  Lynch}]{sung_etal_2012}
Sung, W., M.~S. Ackerman, S.~F. Miller, T.~G. Doak, and M.~Lynch. 2012.
\newblock Drift-barrier hypothesis and mutation-rate evolution.
\newblock Proc Natl Acad Sci USA 109:18489--92.

\bibitem[{Tirosh et~al.(2009)Tirosh, Reikhav, Levy, and
  Barkai}]{tirosh_etal_2009}
Tirosh, I., S.~Reikhav, A.~A. Levy, and N.~Barkai. 2009.
\newblock A yeast hybrid provides insight into the evolution of gene expression
  regulation.
\newblock Science 324:659--62.

\bibitem[{Tirosh et~al.(2010)Tirosh, Sigal, and Barkai}]{tirosh_etal_2010a}
Tirosh, I., N.~Sigal, and N.~Barkai. 2010.
\newblock Divergence of nucleosome positioning between two closely related
  yeast species: genetic basis and functional consequences.
\newblock Mol Syst Biol 6:365.

\bibitem[{Torabi and Kruglyak(2012)}]{Torabi_kruglyak_2012}
Torabi, N., and L.~Kruglyak. 2012.
\newblock Genetic basis of hidden phenotypic variation revealed by increased
  translational readthrough in yeast.
\newblock PLoS Genet 8:e1002546.

\bibitem[{Vakhrusheva et~al.(2011)Vakhrusheva, Kazanov, Mironov, and
  Bazykin}]{vakhrusheva_etal_2011}
Vakhrusheva, A.~A., M.~D. Kazanov, A.~A. Mironov, and G.~A. Bazykin. 2011.
\newblock Evolution of prokaryotic genes by shift of stop codons.
\newblock J. Mol. Evol. 72:138--146.

\bibitem[{Visscher(2008)}]{visscher_2008}
Visscher, P.~M. 2008.
\newblock Sizing up human height variation.
\newblock Nat. Genet. 40:489--490.

\bibitem[{Wagner(2005)}]{wagner_2005}
Wagner, A. 2005.
\newblock Robustness and evolvability in living systems.
\newblock Princeton University Press Princeton, {NJ:}.

\bibitem[{Wagner(2008)}]{wagner_2008}
---{}---{}---. 2008.
\newblock Robustness and evolvability: a paradox resolved.
\newblock Proceedings of the Royal Society B 275:91.

\bibitem[{Wagner(2011)}]{wagner_2011}
---{}---{}---. 2011.
\newblock Genotype networks shed light on evolutionary constraints.
\newblock Trends Ecol Evol 26:577--584.

\bibitem[{Wagner et~al.(2008)Wagner, Kenney-Hunt, Pavlicev, Peck, Waxman, and
  Cheverud}]{wagner_etal_2008}
Wagner, G.~P., J.~P. Kenney-Hunt, M.~Pavlicev, J.~R. Peck, D.~Waxman, and J.~M.
  Cheverud. 2008.
\newblock Pleiotropic scaling of gene effects and the `cost of complexity'.
\newblock Nature 452:470--2.

\bibitem[{Wang et~al.(2010)Wang, Liao, and Zhang}]{wang_etal_2010}
Wang, Z., B.-Y. Liao, and J.~Zhang. 2010.
\newblock Genomic patterns of pleiotropy and the evolution of complexity.
\newblock Proc Natl Acad Sci U S A 107:18034--9.

\bibitem[{Weissman et~al.(2010)Weissman, Feldman, and
  Fisher}]{weissman_etal_2010}
Weissman, D.~B., M.~W. Feldman, and D.~S. Fisher. 2010.
\newblock The rate of fitness-valley crossing in sexual populations.
\newblock Genetics 186:1389.

\bibitem[{Whitehead et~al.(2008)Whitehead, Wilke, Vernazobres, and
  {Bornberg-Bauer}}]{whitehead_etal_2008}
Whitehead, D.~J., C.~O. Wilke, D.~Vernazobres, and E.~{Bornberg-Bauer}. 2008.
\newblock The look-ahead effect of phenotypic mutations.
\newblock Biology Direct 3:18.

\bibitem[{Willi et~al.(2006)Willi, Van~Buskirk, and Hoffmann}]{willi_etal_2006}
Willi, Y., J.~Van~Buskirk, and A.~Hoffmann. 2006.
\newblock Limits to the adaptive potential of small populations.
\newblock Annu. Rev. Ecol. Evol. Syst. 37:433--458.

\end{thebibliography}
\end{document}